\documentclass[twocolumn]{aastex701}

\usepackage{amsmath}

\begin{document}

\title{See you again: an $80^\circ$ NGC~5466 stellar stream and its constraints on the Galactic dark halo mass}

\author[orcid=0009-0007-2466-0416]{Di Bai}
\affiliation{CAS Key Laboratory of Optical Astronomy, National Astronomical Observatories, Chinese Academy of Sciences, Beijing 100101, People's Republic of China}
\affiliation{School of Astronomy and Space Science, University of Chinese Academy of Sciences, No. 19(A) Yuquan Road, Shijingshan District, Beijing, 100049, People's Republic of China}
\email{baidi@bao.ac.cn}

\author[orcid=0000-0001-7609-1947]{Yong Yang}
\affiliation{Institute of Astronomy and Physics, Inner Mongolia University, Hohhot 010021, People's Republic of China}
\email{yangyong@bao.ac.cn}

\author[orcid=0000-0003-2868-8276]{Jing-Kun Zhao} 
\affiliation{CAS Key Laboratory of Optical Astronomy, National Astronomical Observatories, Chinese Academy of Sciences, Beijing 100101, People's Republic of China}
\email{zjk@bao.ac.cn}

\author[orcid=0000-0002-5805-8112]{Xian-hao Ye} 
\affiliation{CAS Key Laboratory of Optical Astronomy, National Astronomical Observatories, Chinese Academy of Sciences, Beijing 100101, People's Republic of China}\affiliation{Institute of Astronomy and Physics, Inner Mongolia University, Hohhot 010021, People's Republic of China}\affiliation{Instituto de Astrof\'{i}sica de Canarias, V\'{i}a L\'{a}ctea, 38205 La Laguna, Tenerife, Spain}\affiliation{Universidad de La Laguna, Departamento de Astrof\'{i}sica, 38206 La Laguna, Tenerife, Spain}
\email{yexianhao@bao.ac.cn}

\author[orcid=0000-0002-8980-945X]{Gang Zhao}
\affiliation{CAS Key Laboratory of Optical Astronomy, National Astronomical Observatories, Chinese Academy of Sciences, Beijing 100101, People's Republic of China}
\affiliation{School of Astronomy and Space Science, University of Chinese Academy of Sciences, No. 19(A) Yuquan Road, Shijingshan District, Beijing, 100049, People's Republic of China}
\email{gzhao@bao.ac.cn}

\correspondingauthor{Yong Yang}
\email[show]{yangyong@bao.ac.cn}
\correspondingauthor{Jing-Kun Zhao}
\email[show]{zjk@bao.ac.cn}

\begin{abstract}

Using \textit{Gaia} Data Release 3 combined with Dark Energy Spectroscopic Instrument Legacy Imaging Survey Data Release 9 and Panoramic Survey Telescope and Rapid Response System Data Release 2, we apply a modified matched filter method to obtain a weighted stellar map in the surrounding of the globular cluster NGC~5466. Our results support that the tidal stream of NGC~5466 extends around $80^\circ$ on the sky, and the leading tail seemingly exhibits an upward deviation in the range of $180^\circ \lesssim \alpha \lesssim 190^\circ$, which supports previous claim that the Large Magellanic Cloud (LMC) may have caused dynamical perturbations to the stream. Based on the 308 highly-weighted member candidates of the NGC~5466 stream, we try to measure the mass distribution of the Milky Way dark matter halo along with the mass of LMC. By comparing the observed stream candidate members to a series of model streams generated under different potential parameters, we obtain the posterior distributions of the parameters: the circular velocity at the solar radius $V_{\rm c} = 239.56^{+1.46}_{-1.44}\ \mathrm{km\ s^{-1}}$, the density flattening of the halo $q = 1.08^{+0.07}_{-0.07}$, the scale radius of the halo density profile $r_s = 12.97^{+3.02}_{-2.18}\ \mathrm{kpc}$ and the total mass of the LMC $M_{\rm LMC} = 9.63^{+1.89}_{-2.00}\times 10^{10}$ $\mathrm{M_\odot}$. The enclosed mass of the Milky Way within the stream apocenter radius of $\sim$ 50 kpc is calculated as $M_{\rm MW}(<50~\mathrm{kpc}) = 4.59^{+0.48}_{-0.37} \times 10^{11}$ $\mathrm{M_\odot}$.

\end{abstract}

\keywords{\uat{Globular star clusters}{656} --- \uat{Stellar streams}{2166} --- \uat{Milky Way dark matter halo}{1049} --- \uat{Large Magellanic Cloud}{903}}


\section{Introduction}
\label{sec:1}

Stellar streams originate from a specific stellar population, such as a globular cluster or a dwarf galaxy, typically appearing as narrow, elongated structures in the Milky Way halo \citep[see e.g.,][]{2023MNRAS.520.5225M,2025NewAR.10001713B,2025AJ....170..157K}. Dynamically, streams are fragile such that interactions with massive perturbers can alter the orbits of stars in them, making streams powerful tracers of the gravity of these perturbers \citep{2002ApJ...570..656J,2002MNRAS.332..915I}. On the one hand, small-scale features in streams can arise from close flybys with such as invisible dark matter subhalos \citep[e.g.,][]{2015MNRAS.450.1136E,2015MNRAS.454.3542E,2019ApJ...880...38B,2021MNRAS.502.2364B,2022ApJ...941..129D,2025ApJ...989L..52Z}. On the other hand, global orbital trajectories of streams are highly sensitive to the gravitational potential,  since variations in the Galactic mass distribution and potential shape can significantly influence their morphology and orbital trajectories. Therefore, stellar streams have been widely used to constrain the mass distribution and structure of the Milky Way. For example, \citet{2015AAS...22514219K} used the tidal tails of the globular cluster Palomar 5 to constrain the mass and flattening of the dark matter halo, while \citet{2016ApJ...833...31B} measured the halo mass using the Palomar 5 and GD-1 streams in action space. \citet{2019MNRAS.486.2995M} fitted the GD-1 orbit to constrain the circular velocity at the solar radius and the halo shape, \citet{2023MNRAS.521.4936K} used the Orphan--Chenab stream to constrain the mass profile of the Milky Way, and \citet{2025ApJ...985L..22N} used accelerations along the GD-1 stream to constrain the mass and flattening of the dark matter halo. These studies demonstrate that the spatial and kinematic information encoded in stellar streams provides a powerful approach for probing the Galactic gravitational potential.

NGC~5466 is a globular cluster embedded in the Galactic halo, whose tidal tails represent a typical cold stellar stream. \citet{2006ApJ...637L..29B} used a neural network to identify cluster member stars and discovered a stellar stream of roughly $4^\circ$ in length. \citet{2006ApJ...639L..17G} applied a matched filter technique to Sloan Digital Sky Survey photometric data and detected the NGC~5466 stream extending over $45^\circ$ in right ascension ($\alpha$), from $\alpha \approx 180^\circ$ to $225^\circ$. They also noted that in the region $\alpha < 190^\circ$, the stream signal deviated from the cluster's orbit, suggesting that this ``offset'' was likely caused by interactions with other massive structures. \citet{2007MNRAS.380..749F} simulated the NGC~5466 stream and obtained a length exceeding $100^\circ$, which partially supported the $45^\circ$ observational result. Using the \texttt{STREAMFINDER} algorithm \citep{2018MNRAS.477.4063M}, \citet{2021ApJ...914..123I} were the first to identify individual member stars in the NGC~5466 stream. \citet{2021MNRAS.507.1923J} applied constraints in phase space and metallicity to obtain 33 member stars spanning approximately $31^\circ$ and found good agreement with a mock stream. They also suggested that the signal at $\alpha < 200^\circ$ reported by \citet{2006ApJ...639L..17G} may not correspond to the genuine stream, as no members were detected within that region and their mock stream deviated from the matched filter result of \citet{2006ApJ...639L..17G}. \citet{2022MNRAS.513..853Y} identified 11 spectroscopic member stars spanning $60^\circ$ of the sky and simulated the disruption of NGC~5466 in a gravitational potential composed of the Milky Way and the Large Magellanic Cloud (LMC). The resulting model showed a good agreement with the stream members, indicating that LMC may have induced dynamical perturbations on the NGC~5466 leading tail, which was also presented in \citet{2022MNRAS.510.2437E}. \citet{2024ApJ...967...89I} released the latest \texttt{STREAMFINDER} detections of 87 streams, including a $\sim 23^\circ$ long tail around NGC~5466, which implied that the stream at far sides from the cluster might be quite dim and weak.

In terms of dynamics of the NGC 5466 stream, \citet{2012MNRAS.424L..16L} argued that the offset region ($\alpha < 190^\circ$) provides strong constraints on the shape of the Galactic dark matter halo, and further maps of the westward end of the stream as well as radial velocities/distances
along the whole known stream are the key to unravelling the MW halo shape
using the NGC 5466 stream. \citet{2013MNRAS.436.2386L} pointed out that the tidal tails of
the globular clusters NGC 5466 and Palomar 5 are promising for
constraining the Galactic potential. \citet{2021MNRAS.507.1923J} highlighted
that NGC 5466 serves as a unique test bed for dynamical modeling of the
Milky Way, owing to its large Galactocentric distance, pronounced distance
gradient, known progenitor cluster, and recent interaction with the Galactic
disk. Further emphasizing its scientific importance, NGC~5466 stream is included in the planned Subaru \textquoteleft Ōnohi\textquoteleft ula Prime Focus Spectrograph (PFS) Outer Milky Way fields \citep{2026arXiv260409875C}. Importantly, NGC 5466 may also provide a useful probe of the dynamical influence of the Large Magellanic Cloud (LMC). Numerical simulations by \citet{2022MNRAS.510.2437E} showed that a close encounter between NGC~5466 and the LMC can produce prominent morphological features in the tidal tails of NGC~5466, including a spur-like feature for tail stars with $x$-coordinates greater than that of the cluster and a kink for tail stars with $x$-coordinates less than that of the cluster.
 In their simulations, NGC 5466 passed within approximately 24~kpc of the LMC at a relative velocity of 376.5~km~s$^{-1}$ about 160~Myr ago. Similarly, \citet{2022MNRAS.513..853Y} investigated the NGC 5466 stream in a Milky Way potential including a moving LMC and found that the LMC perturbation can significantly affect the morphology and kinematics of the modeled stream. These results indicate that the sensitivity of NGC 5466 to the LMC is determined not simply by its present-day location on the sky, but by its orbital history and the long-term dynamical interaction between the LMC and the extended tidal stream.

In summary, our knowledge of the NGC~5466 stellar stream is still far from complete. 
Several important questions remain unresolved. First, the full
extent and morphology of the stream are still uncertain. Although previous
studies have reported an extended structure spanning several tens of degrees,
the detailed continuity of the stream, particularly toward its western and
more distant regions, remains to be firmly established. Second, the origin
of the apparent deviation of the stream from the progenitor's orbital path
remains unclear. Third, the role of the Large Magellanic Cloud (LMC) in shaping the NGC 5466 stream has not yet been quantitatively established. Numerical simulations have shown that a close encounter with the LMC can produce prominent morphological features, including a kink and a spur in the tidal tails \citep{2022MNRAS.510.2437E}, and \citet{2022MNRAS.513..853Y} further demonstrated that including the LMC can affect the modeled stream morphology and kinematics. However, these studies adopted a fixed LMC mass rather than determining it directly from the NGC 5466 stream. Thus, an important open
question is whether the observed morphology and phase-space structure of
NGC 5466 contain sufficient information to quantitatively constrain the LMC
mass. Addressing these issues requires a more complete and homogeneous sample
of stream members, together with a dynamical model that simultaneously
accounts for the Milky Way potential and the perturbation induced by the LMC.

In this work, we keep focusing on the NGC 5466 stream, including its detection and dynamic applications. The paper is organized as follows. Section~\ref{sec:2} describes the photometric and astrometric data used in this work. Section~\ref{sec:3} constructs the weighted stellar density map using a modified matched filter technique to identify the extended tidal tails of NGC~5466, and presents a stream member candidate catalog. In Section~\ref{sec:4}, we constrain the Galactic dark matter halo and the mass of the LMC. In Section~\ref{sec:5}, we discuss our results in context of previous literature. Conclusions are summarized in Section~\ref{sec:6}.

\section{Data}
\label{sec:2}

We detect NGC~5466 stream by using both astrometry and photometry. Astrometry comes from \textit{Gaia} Data Release 3 \citep[\textit{Gaia} DR3][]{2023A&A...674A...1G}. Although \textit{Gaia} DR3 provides photometry as well, we replace it with Dark Energy Spectroscopic Instrument Legacy Imaging Survey Data Release 9 \citep[LS DR9][]{2019AJ....157..168D} combined with Panoramic Survey Telescope and Rapid Response System Data Release 2 \citep[PS DR2][]{2020ApJS..251....6M}.  Because LS DR9 and PS DR2 reach much deeper limiting magnitudes than \textit{Gaia}, this hybrid approach ensures high photometric precision near the \textit{Gaia} detection limit ($G \sim 21\,\mathrm{mag}$). We describe details of data processing in the following.

\subsection{Gaia DR3}

In order to guarantee good astrometry, we require \textit{Gaia} DR3 sources to have the renormalized unit weight error \texttt{ruwe} $<1.4$ \citep{2021A&A...649A...2L}. The parallax $\varpi$ zero-point is corrected based on codes from \citet{2021A&A...649A...4L}. Given that the distance of the NGC~5466 cluster is $\sim 16$~kpc, we remove all foreground stars within a distance of 10~kpc from the Sun, using a conservative $3\sigma$ confidence interval with the selection
\begin{equation}
\varpi - 3\sigma_\varpi < 0.1 \ \mathrm{mas}.
\end{equation}

\subsection{LS DR9}

We download LS DR9 in the $g$, $r$, and $z$ bands that overlap with the model on the sky, from right ascension $\alpha$ and declination $\delta$ at ($170^\circ$,~$+50^\circ$) to around ($260^\circ$,~$0^\circ$), where the latter end corresponds to the north footprint boundary of LS DR9. Similar to \citet{2019ApJS..242....8Z}, we apply following data quality cuts
\begin{equation}
\begin{aligned}
    \mathrm{TYPE} &= \mathrm{PSF} , \\
    \mathrm{FLUX\_G, R, Z} &> 0 , \\
    \mathrm{ALLMASK\_G, R, Z} &= 0 , \\
    \mathrm{FRACMASKED\_G, R, Z} &< 0.5 , \\
    \mathrm{FRACFLUX\_G, R, Z} &< 0.5 , \\
    \mathrm{FRACIN\_G, R, Z} &> 0.3   .
\end{aligned}
\end{equation}
Here, ``ALLMASK'' and “FRACMASKED” are used to remove sources with masked pixels. ``FRACFLUX'' and ``FRACIN'' are used to avoid serious flux contamination from nearby objects \citep{2019ApJS..242....8Z}. Extinction-corrected magnitudes are obtained with
\begin{equation}
    \mathrm{g_0,r_0,z_0} = 22.5 - 2.5 \mathrm{log}_{10}
    \left( \frac{\mathrm{FLUX\_G,R,Z}}{\mathrm{MW\_TRANSMISSION\_G,R,Z}} \right) .
\end{equation}
We further calculate magnitude errors using error propagation
\begin{equation}
    \sigma_\mathrm{g,r,z} = \frac{2.5}{\mathrm{ln}(10)}
    \frac{1}{ \mathrm{FLUX\_G,R,Z} \times \sqrt{\mathrm{FLUX\_IVAR\_G,R,Z}} } .
\end{equation}

\subsection{PS DR2}

Due to the coverage of LS DR9, we complement photometry with PS DR2. Specifically, we fetch PS DR2 in the $g$, $r$, and $z$ bands beyond ($\alpha$,~$\delta$) = ($260^\circ$,~$0^\circ$) above the Galactic plane, galactic longitude $l$ and latitude $b$ between $5^\circ < l < 35^\circ$ and $15^\circ < b < 30^\circ$. We require
\begin{equation}
\begin{aligned}
    \mathrm{g,r,zMeanPSFMag} &> 0 , \\
    \mathrm{g,r,zMeanPSFMagErr} &> 0 .
\end{aligned}
\end{equation}
We then correct for interstellar extinctions using the dust map from \citet{1998ApJ...500..525S} and dereddening coefficients from \citet{2011ApJ...737..103S}. 

After the processing above, we combine LS DR9 and PS DR2 together, and then cross match with \textit{Gaia} DR3, as the final data set. The precise astrometric measurements provided by \textit{Gaia} significantly reduce contamination from extended non-stellar sources. Furthermore, for LS DR9 photometry, we require $\mathrm{TYPE}=\mathrm{PSF}$ during the quality selection process, which further favors point-like sources and minimizes residual galaxy contamination.

\begin{figure*}[t]
\centering
{\bf LS DR9}

\includegraphics[width=0.8\textwidth]{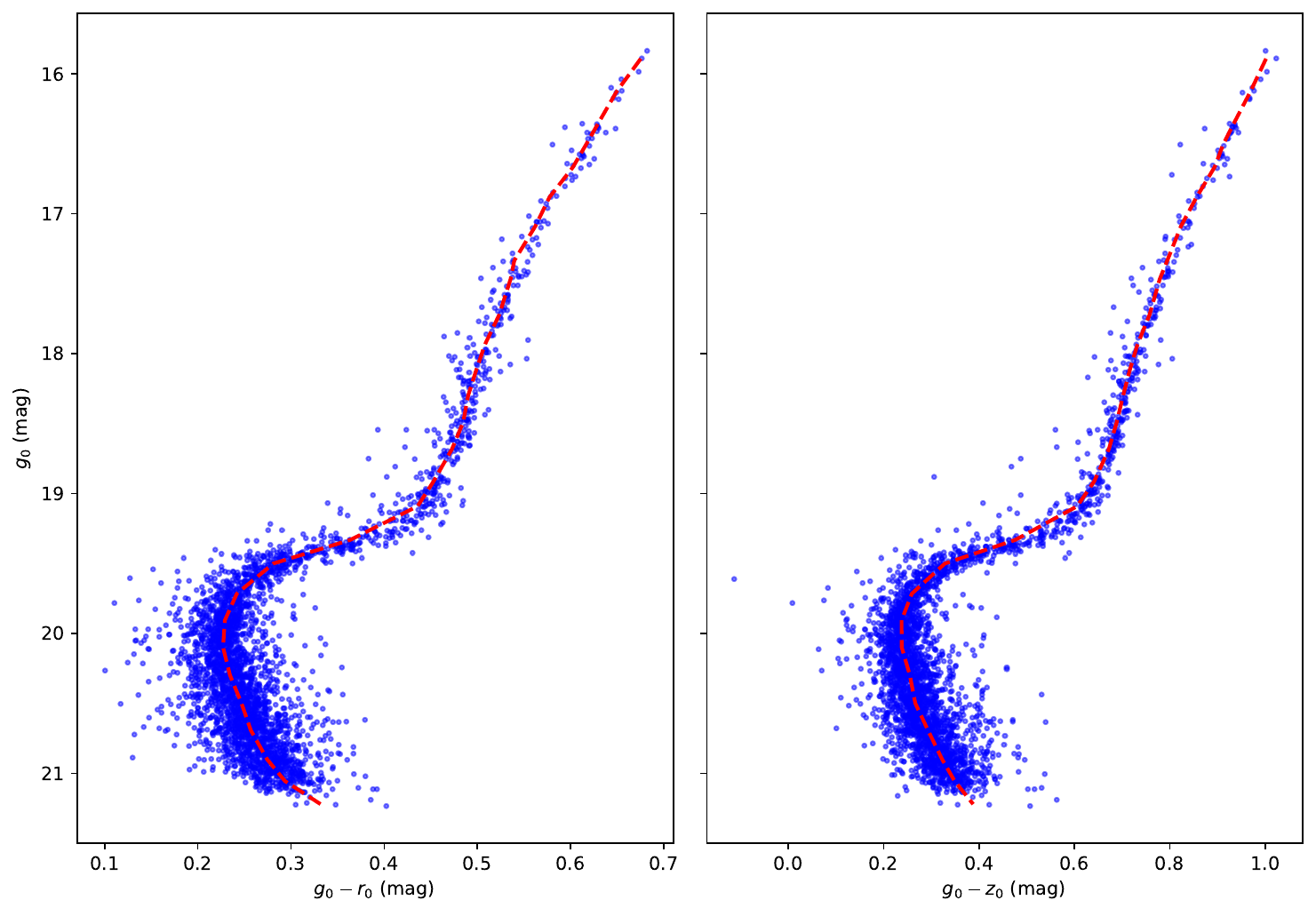}

{\bf PS DR2}

\includegraphics[width=0.8\textwidth]{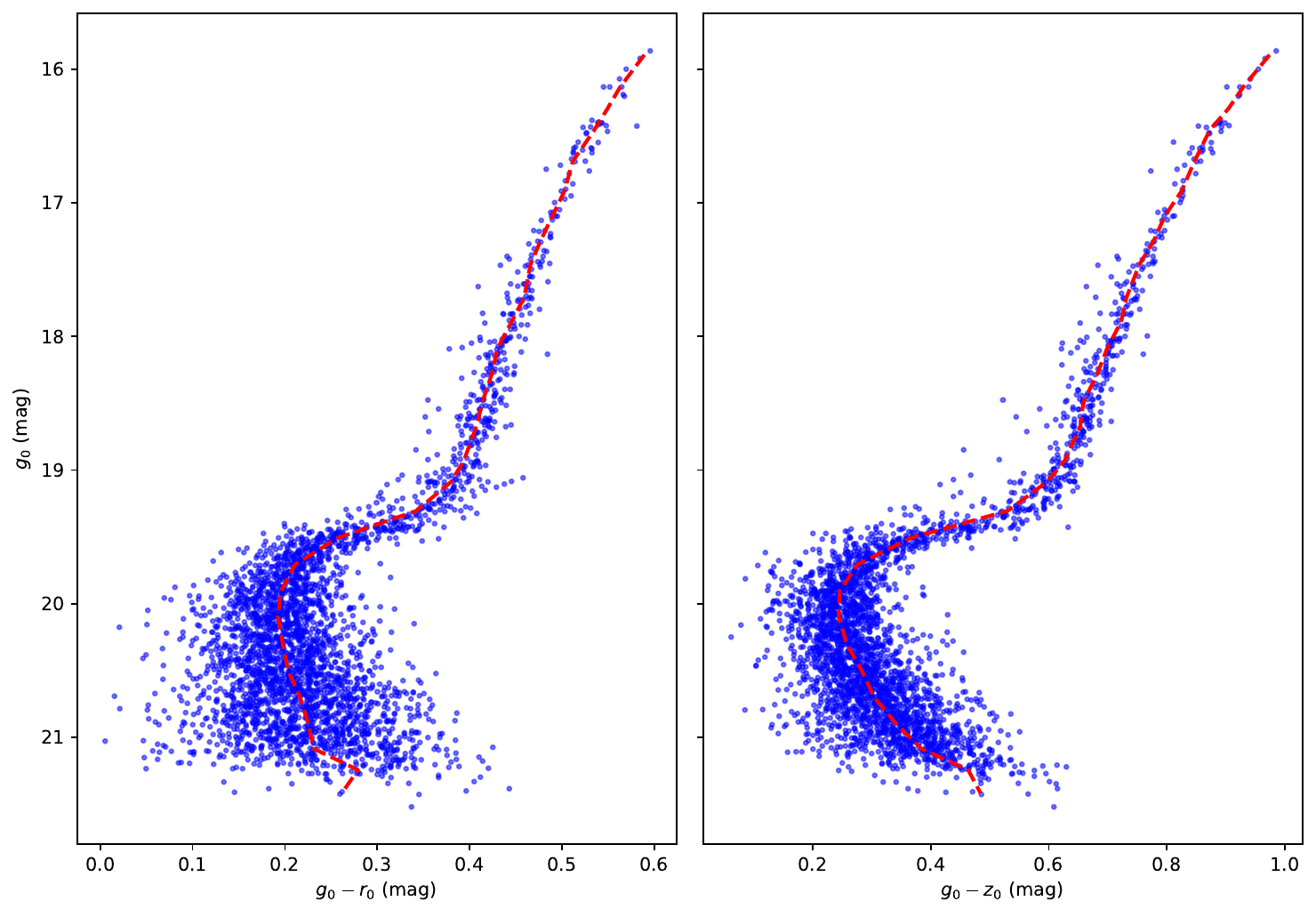}
\caption{
CMDs of stars within the tidal radius of NGC~5466. 
The upper and lower panels show CMDs derived from LS and PS, respectively. 
Blue dots represent individual stars, and the red dashed line traces the locus of the red giant branch and main sequence.
}
\label{fig:1}
\end{figure*}

\begin{figure*}[ht!]
\centering
\includegraphics[width=0.7\textwidth]{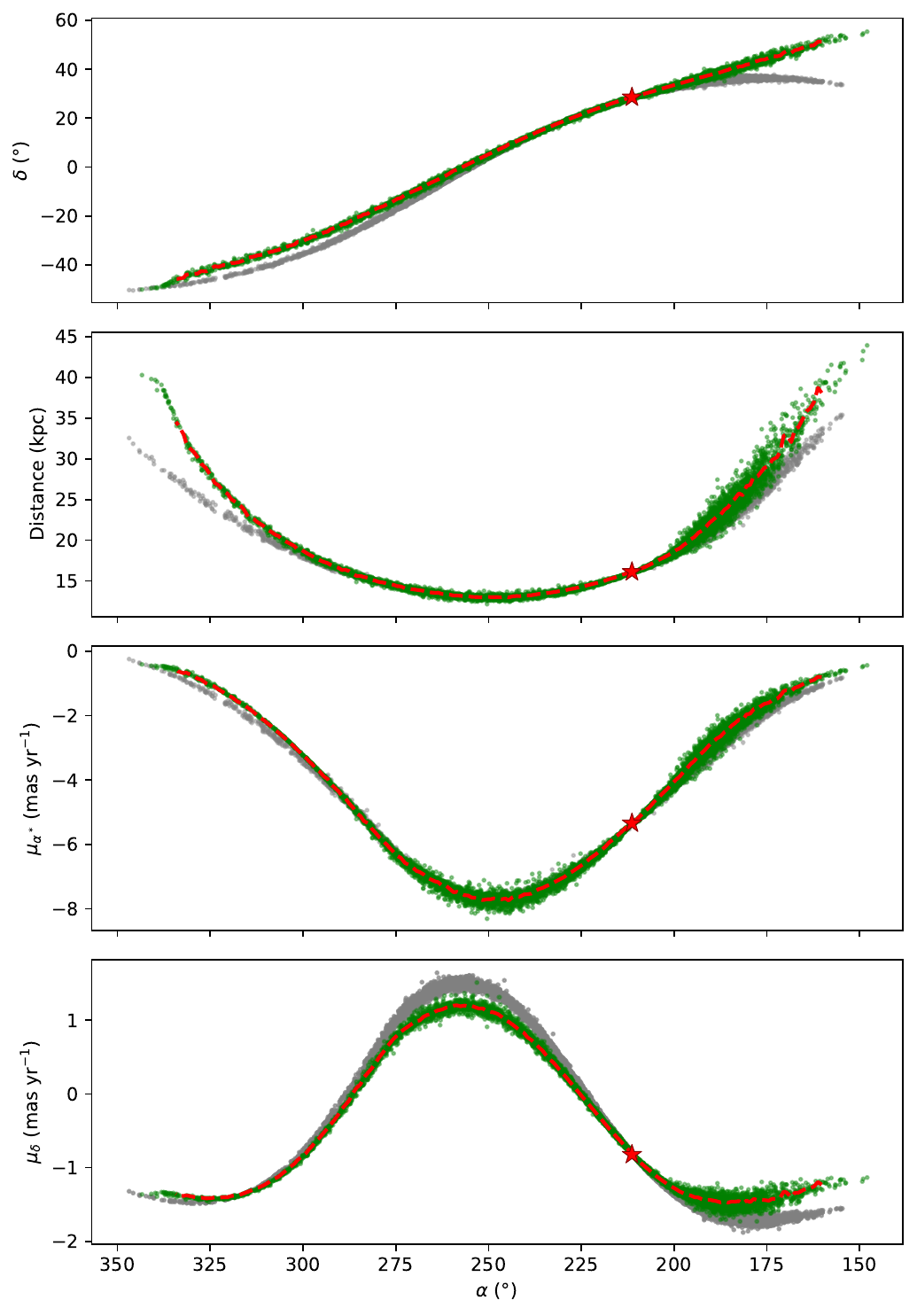}
\caption{Green dots show the mock stream obtained with the gravitational influence of the LMC included in planes of $\delta$, heliocentric distance, $\mu_\alpha^*$, and $\mu_\delta$ as a function of $\alpha$. The red dashed lines indicate the computed medians along $\alpha$. Grey dots show the mock stream obtained without including the gravitational influence of the LMC. The red star marks the location of the present-day values for NGC~5466.
}
\label{fig:3}
\end{figure*}

\section{Detection of NGC~5466 stream}
\label{sec:3}
In this section, we identify member candidates of the NGC~5466 tidal stream using the data obtained in Section~\ref{sec:2}.

\subsection{Modified Matched Filter}
\label{sec:3.1}

Prior to the \textit{Gaia} era, the detection of stellar streams primarily relied on the application of matched filters \citep{2002AJ....124..349R} in a colour–magnitude diagram (CMD), under the assumption that streams’ stellar populations follow specific isochrones \citep[e.g.,][]{2006ApJ...643L..17G}. The arrival of \textit{Gaia} makes it feasible to further incorporate proper motion filters to enhance the contrast between streams and foreground stars \citep[e.g.,][]{2019ApJ...884..174G,2023ApJ...945L...5Y,2026MNRAS.547ag488Y}. Following the approach of \citet{2019ApJ...884..174G}, we adopt a modified matched filter to detect the stellar stream of NGC~5466 by combining photometric and astrometric information. The modified filter assigns weights to stars assuming a Gaussian error distribution in the CMD space and the proper motion (PM) space, taking photometric and astrometric uncertainties into account.

We construct a CMD-filter for NGC~5466 using stars located within the cluster's tidal radius. We adopt a tidal radius of $r_{\rm t}=15.68'$ for NGC~5466, taken from the \citet{1996AJ....112.1487H} catalog. For the red giant branch and main sequence stars, we calculate the median color along the extinction-corrected $g_0$ magnitude to define the CMD locus.
This locus serves as the reference for assigning the CMD weights $w_{\rm CMD}$ to individual stars:
\begin{equation}
w_{\rm CMD} = \frac{1}{\sqrt{2 \pi} \sigma_{\rm color}} \exp \left[ -\frac{1}{2} \left( \frac{{\rm color} - {\rm color}_0}{\sigma_{\rm color}} \right)^2 \right],
\end{equation}
where \text{color} is the measured $g_0-r_0$ and $g_0-z_0$ of a star, and $\sigma_{\rm color}$ is its uncertainty, and ${\rm color}_0$ is derived from the CMD-filter at the given $g_0$ magnitude of the star. We compute CMD-based weights separately for $g_0-r_0$ and $g_0-z_0$, and take their product as the final weight. Figure \ref{fig:1} shows the CMDs of stars within the tidal radius of NGC~5466. The upper and lower panels show CMDs derived from LS DR9 and PS DR2, respectively. The blue dots represent individual stars, and the red dashed line traces the locus of the red giant branch and main sequence.

\begin{table}[htbp]
\centering
\caption{Adopted present-day phase-space coordinates and structural parameters of NGC~5466 used in the particle-spray stream modeling.}
\label{tab:1}
\begin{tabular}{lcc}
\hline\hline
Parameter & Symbol & Adopted Value \\
\hline
Right ascension & $\alpha$ & $211.363709^\circ$ \\
Declination & $\delta$ & $28.534445^\circ$ \\
Heliocentric distance & $d$ & $16.12\,\mathrm{kpc}$ \\
Proper motion in Right ascension & $\mu_{\alpha^*}$ & $-5.336\,\mathrm{mas\,yr^{-1}}$ \\
Proper motion in Declination & $\mu_{\delta}$ & $-0.809\,\mathrm{mas\,yr^{-1}}$ \\
Radial velocity & $v_{\rm los}$ & $106.82\,\mathrm{km\,s^{-1}}$ \\
Progenitor mass & $M_{\rm prog}$ & $6\times10^{4}\,M_\odot$ \\
Plummer scale radius & $r_{\rm scale}$ & $13\,\mathrm{pc}$ \\
\hline
\end{tabular}
\end{table}

We further obtain expected PMs along the tidal tails by generating a mock stream for NGC~5466 using the improved particle spray algorithm from \citet{2025ApJS..276...32C}. Briefly, star particles can be released from the Lagrange points of the progenitor while orbiting from the past to present day. These particles follow certain phase space distributions relative to the cluster when they are released, after which they gradually move far away to form the tidal tails. The gravitational potential of the NGC~5466 is modeled using a Plummer sphere \citep{1911MNRAS..71..460P} with a mass and scale radius of $6\times10^4$ M$_\odot$ and 13 pc \citep{2018MNRAS.478.1520B}. The cluster's location and velocity are taken from \citet{2021MNRAS.505.5978V} and \citet{2021MNRAS.505.5957B}. The above information is summarized in Table~\ref{tab:1}. For the host gravitational potential, we adopt the \texttt{MWPotential2014} in \texttt{galpy} package \citep{2014ascl.soft11008B}. Even though the NGC~5466 stream is located in the north Galaxy, both \citet{2022MNRAS.510.2437E} and \citet{2022MNRAS.513..853Y} showed that including LMC could significantly alter the leading tail of the cluster. Therefore, we add a moving LMC in the potential, which is modeled by a truncated NFW profile \citep{1997ApJ...490..493N} with a mass of $1.5\times10^{11}$ M$_\odot$ \citep{2019MNRAS.487.2685E,2023MNRAS.521.4936K} and a scale radius of 10.84 $\mathrm{kpc}$, which is derived by
\begin{equation}
r_\mathrm{scale}=8.5\ \mathrm{kpc} \times ( M_{\rm LMC}/10^{11}\ \mathrm{M_\odot} )^{0.6} .
\label{eq:7}
\end{equation}
The time-dependent potential also takes into account the LMC’s gravitational pull on the Milky Way, which causes the latter to be a non-inertial reference frame. These LMC-related settings (including Equation~\ref{eq:7}) are all adopted from \citet{2021MNRAS.501.2279V}. We adopt the solar Galactocentric distance $R_0=8.122\,{\rm kpc}$ \citep{2018A&A...615L..15G}. Here we adopt attributes of ``v4.0'' in \texttt{astropy} \citep{2022ApJ...935..167A} to transform coordinates between ICRS and Galactocentric frames.

\begin{figure*}[t]
\centering
\includegraphics[width=0.8\textwidth]{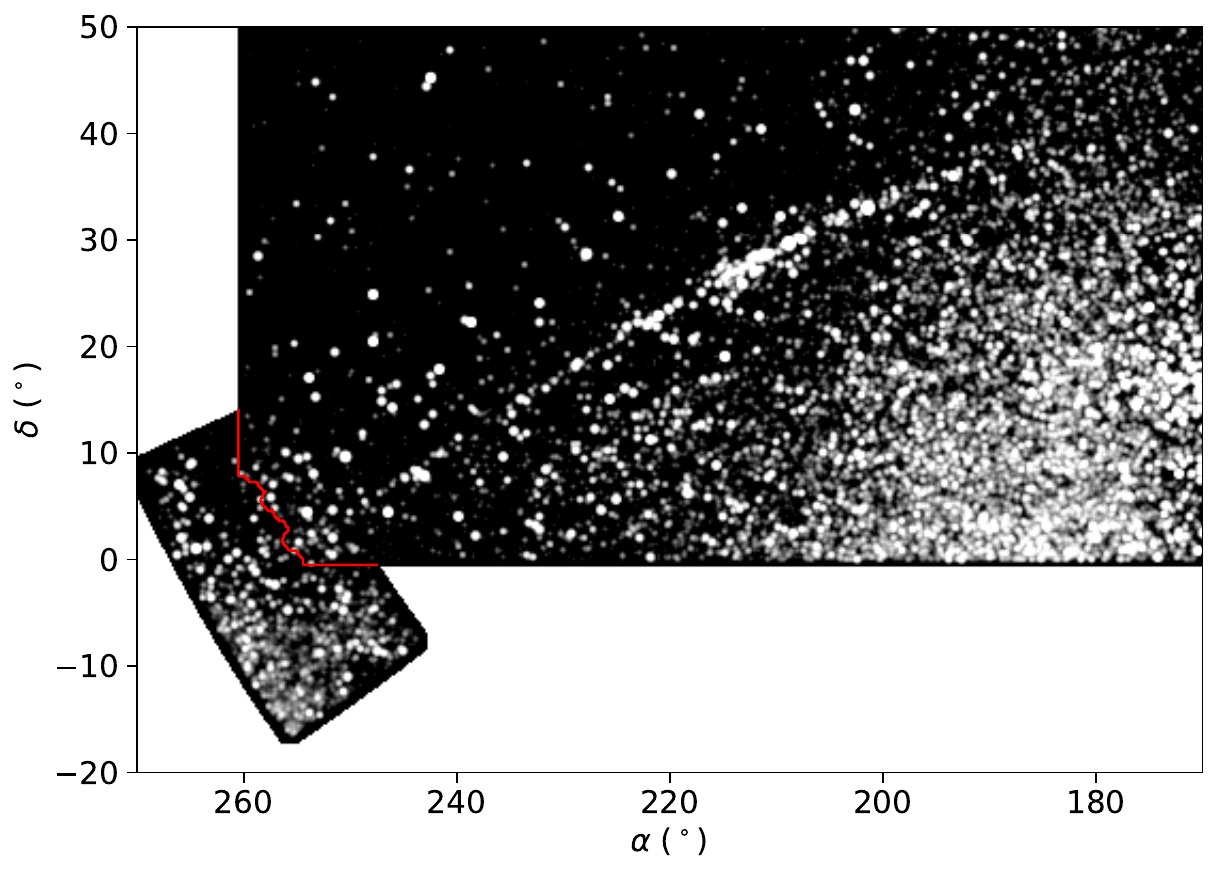}
\includegraphics[width=0.8\textwidth]{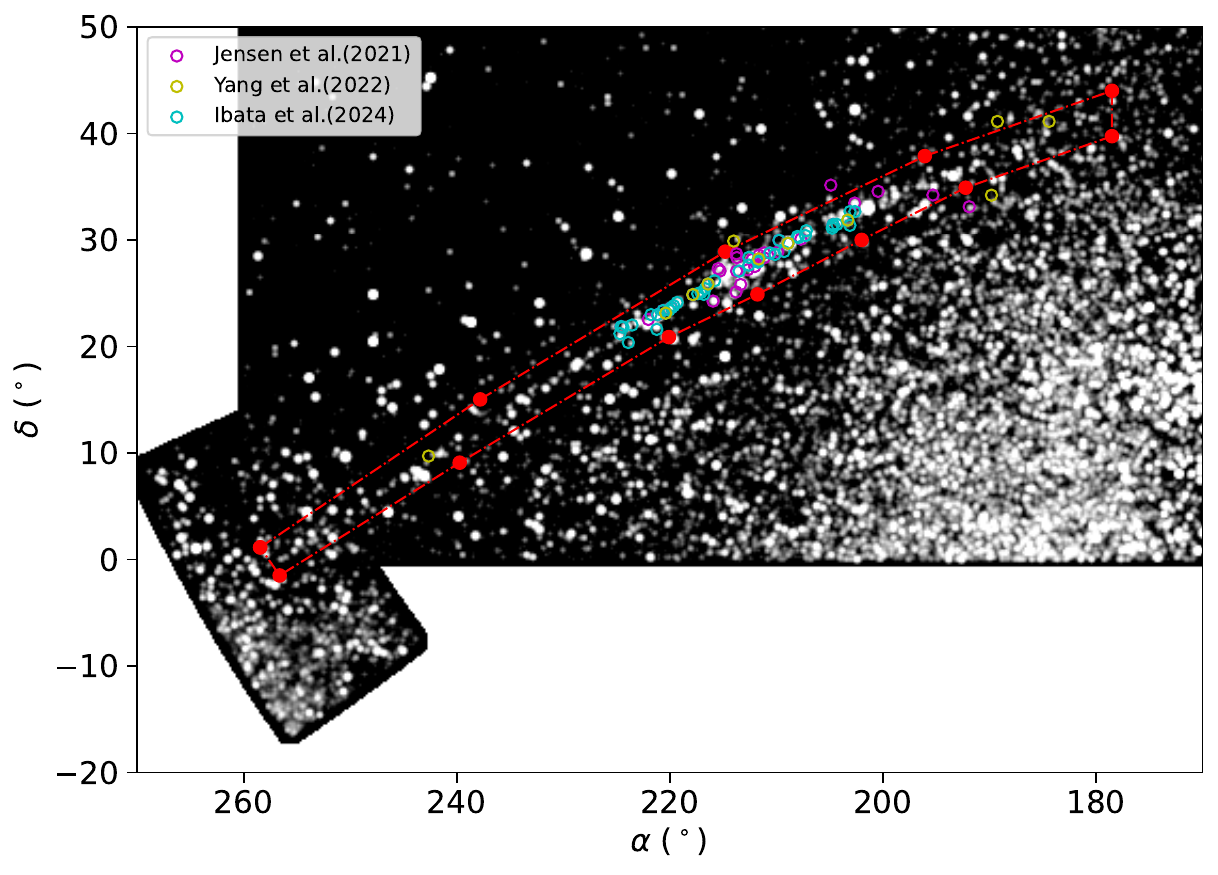}
\caption{Upper panel: weighted stellar density map in the region surrounding the NGC~5466 globular cluster. Stellar weights are summed in sky pixels of $0.2^\circ \times 0.2^\circ$, and the map is smoothed with a Gaussian kernel of $\sigma = 0.8$ pixels. The region to the right of the red line is covered by LS DR9 data, while the region to the left is covered by PS DR2 data. Lower panel: spatial distribution of member stars identified in three previous studies. The red dot-dashed lines indicate the predefined sky region adopted for member selection, based on the spatial extent of the NGC~5466 stream.}
\label{fig:4}
\end{figure*}

To obtain a tidal stream with spatial extent comparable to the observational footprint, the orbit of NGC~5466 is first integrated backward in time for 5 Gyr and then evolved forward to the present day. At intervals of 1~Myr, star particles are released from the Lagrange points of the cluster, following the procedure of \citet{2025ApJS..276...32C}. This is achieved with \texttt{AGAMA} \citep{2019MNRAS.482.1525V}. Figure \ref{fig:3} shows the mock stream projected in $\delta$, heliocentric distance, $\mu_{\alpha}^{*}$, and $\mu_{\delta}$ as functions of $\alpha$. The red dashed curves indicate the corresponding median trends. The green dots represent the mock stream with the gravitational influence of the LMC, while the grey dots represent the mock stream without including the gravitational influence of the LMC as a comparison. Notably, a clear deviation from the main stream track is observed in the range of $\alpha \lesssim 190^\circ$, where the leading tail exhibits an upward bend. Such a morphological distortion is suggestive of a dynamical perturbation, potentially induced by the gravitational influence of the LMC. In our fiducial particle-spray model, the present-day distance, proper motions, and radial velocity of NGC~5466 are fixed to their nominal observational values. To assess whether this assumption could introduce a significant variation in the predicted stream trajectory, we performed an additional test in which the progenitor's phase-space coordinates were varied within their observational uncertainties and the corresponding stream realizations were compared with the fiducial model. We find that the resulting variation in the stream morphology and kinematics is small. The difference in the perpendicular stream width is only $0.0132^\circ$, corresponding to $0.79 \ \mathrm{arcmin}$ and a relative difference of $4.79\%$.
The difference in the standard-deviation estimate of the perpendicular width is $0.0301^\circ$, corresponding to $1.81 \ \mathrm{arcmin}$ and a relative difference of $7.09\%$.
The proper-motion dispersions differ by only $0.007~\mathrm{mas\ yr^{-1}}$ ($5.65\%$) in $\sigma_{\mu_{\alpha*}}$ and $0.002~\mathrm{mas\ yr^{-1}}$ ($4.26\%$) in $\sigma_{\mu_\delta}$.

For each star, the PM weight $w_{\rm PM}$ is computed by evaluating its distance from the PM-filters in the two-dimensional proper-motion space. Assuming Gaussian uncertainties, the PM weight is defined as

\begin{multline}
w_{\rm PMs} =
\frac{1}{2\pi\,\sigma_{\mu_{\alpha^*}}\,\sigma_{\mu_\delta}}
\exp\Bigg[
-\frac{1}{2}
\Bigg(\frac{\mu_{\alpha^*}-\mu_{\alpha,0}^*}
{\sigma_{\mu_{\alpha^*}}}\Bigg)^2 \\
-\frac{1}{2}
\Bigg(\frac{\mu_{\delta}-\mu_{\delta,0}}
{\sigma_{\mu_\delta}}\Bigg)^2
\Bigg],
\end{multline}
where $(\mu_{\alpha^*}, \mu_\delta)$ are the observed proper motion components of a star, $(\sigma_{\mu_{\alpha^*}}, \sigma_{\mu_\delta})$ are the corresponding observational uncertainties, and $(\mu_{\alpha^*,0}, \mu_{\delta,0})$ are the median proper motions predicted by the mock stream at the star's $\alpha$. This formulation assigns higher weights to stars whose proper motions closely follow the mock stream in both PM dimensions. Notably, for data at different $\alpha$ segments, the CMD filters are shifted at appropriate distance moduli as predicted by the model stream.

The final weight for each star is obtained by multiplying the CMD and PM weights, $w_{\rm CMD} \times w_{\rm PMs}$. 
Within each sky pixel, the weights of all stars are summed to produce a weighted density map, from which potential stream structures can be identified.

\subsection{Sky Map and Stream Identification}

Using the weights of all stars, we construct a weighted stellar density map in the region surrounding the NGC~5466 globular cluster, which is shown in the upper panel of Figure~\ref{fig:4}. Stellar weights are summed in sky pixels of $0.2^\circ \times 0.2^\circ$, and the map is smoothed with a Gaussian kernel of $\sigma = 0.8$ pixels. The lower panel of Figure~\ref{fig:4} shows the spatial distribution of member stars identified in previous studies \citep[][]{2021MNRAS.507.1923J,2022MNRAS.513..853Y,2024ApJ...967...89I}. The red dot-dashed lines indicate the predefined sky region adopted for member selection, based on the spatial extent of the NGC~5466 stream. An inspection of the weighted sky map reveals that the tidal stream of NGC~5466 extends around $80^\circ$ on the sky, from $\alpha=180^\circ$ to $260^\circ$. The stream can be continuously traced across this angular range, indicating that the tidal debris of NGC~5466 is more extended than previously reported. We generated an independent particle-spray stream simulation using a particle-spray stream generated without the LMC and recomputed the weighted stellar density map across the full survey footprint. A direct comparison between the LMC-free density map and our fiducial map reveals the spatial trajectory, high-density stream spine, and overall morphology of the NGC~5466 stream are essentially identical in both maps. Based on the ranking of their weights, a total of 343 stars with weights exceeding 20 are identified as member candidates within the red polygon. A threshold of $w_{\rm cut}=20$ is chosen as it represents the optimal balance between completeness and purity: a lower cut allows significant contamination from background stars to mimic spurious stream-like features, while a higher cut compromises completeness by discarding bona fide stream members and truncating the visible stream length. The weight distribution is highly skewed: among the stars outside the NGC~5466 tidal radius, more than $99.9\%$ have $w<0.72$, while only $0.0238\%$ satisfy the adopted threshold of $w>20$. Within the tidal radius of NGC~5466, stars associated with the progenitor region show significantly higher weights, with a median value of $w=3.45$ and an upper quartile of $Q_3=56.67$. The stars selected with $w>20$ have a median weight of $w=128.14$ (interquartile range: $51.75$--$485.27$). To minimize model dependence and improve the robustness of the member candidate selection with respect to the assumed Milky Way potential, we independently calculate the weights using both \texttt{MilkyWayPotential2022} from \texttt{gala} package \citep{2017JOSS....2..388P} and \texttt{MWPotentialHunter24} from \texttt{galpy} package \citep{2014ascl.soft11008B}. We apply the same selection threshold of $w>20$ to both sets of weights. The \texttt{MilkyWayPotential2022} and \texttt{MWPotentialHunter24} models both yield 342 member candidates, with 308 stars in common among all three potentials. We adopt these 308 common stars as the member candidates within the red polygon. Table~\ref{tab:mw2022} presents the model parameters adopted from \texttt{MWPotential2014}. Table~\ref{tab:2} provides the catalog of member candidates in the stellar stream. Compared with traditional ``brute-force'' filtering methods based on fixed cuts in sky position, proper motion, and color--magnitude space, the modified matched-filter technique provides a continuous weighting scheme that quantitatively evaluates the consistency of individual stars with the expected stream properties. It is worth noting that the modified matched-filter technique suppresses contamination from the Milky Way foreground through the combined CMD and proper-motion weighting.

\begin{table*}[htbp]
\centering
\caption{
Model Parameters Adopted from \texttt{MWPotential2014}}
\label{tab:potential_mcmc_priors}

\begin{tabular}{lllc}
\hline\hline

Component & Parameter & Adopted Value \\

\hline

Bulge 
& Profile Type 
& Spheroid 
\\

& Density normalization 
& $2.22694\times10^{8}\,
\mathrm{M_\odot\,kpc^{-3}}$
\\

& Inner and outer slope 
& $1.8,\ 1.8$
\\

& Scale radius 
& $1.0\,\mathrm{kpc}$
\\

& Cutoff radius 
& $1.9\,\mathrm{kpc}$
\\

\hline

Disk 
& Profile Type 
& Miyamoto--Nagai
\\

& Radial scale length 
& $3.0\,\mathrm{kpc}$
\\

& Vertical scale length 
& $0.28\,\mathrm{kpc}$
\\

& Disk mass 
& $6.819386\times10^{10}\,\mathrm{M_\odot}$
\\
\hline
Dark Matter Halo & Profile Type & NFW (Spheroid) \\ & Density normalization & $1.273\times10^{7}\, \mathrm{M_\odot\,kpc^{-3}}$ \\ & Inner and outer slope & $1.0,\ 3.0$ \\ & Scale radius & $16.0\,\mathrm{kpc}$ \\ \hline

\label{tab:mw2022}
\end{tabular}
\end{table*}

\begin{deluxetable*}{lcccccccccc}
\tablecaption{Selected high-weight member candidates of the NGC~5466 stellar stream (first five entries). This table is available in its entirety in machine-readable form in the online article. \label{tab:2}}
\tablewidth{\textwidth}
\tabletypesize{\footnotesize} % 如果页面依然略紧，可用 footnotesize 缩微
\tablehead{
\colhead{Source ID} & \colhead{$\alpha$} & \colhead{$\delta$} & \colhead{$\mu_{\alpha^*}$} & \colhead{$\sigma_{\mu_{\alpha^*}}$} & \colhead{$\mu_\delta$} & \colhead{$\sigma_{\mu_\delta}$} & \colhead{$g_0$} & \colhead{$r_0$} & \colhead{$z_0$} & \colhead{w throughout paper} \\
\colhead{} & \colhead{(deg)} & \colhead{(deg)} & \colhead{(mas~yr$^{-1}$)} & \colhead{(mas~yr$^{-1}$)} & \colhead{(mas~yr$^{-1}$)} & \colhead{(mas~yr$^{-1}$)} & \colhead{} & \colhead{} & \colhead{} & \colhead{}
}
\startdata
1533307997596795392 & 184.92 & 40.54 & $-1.29$ & 0.72 & $-1.64$ & 0.80 & 20.59 & 20.35 & 20.33 & 20.32 \\
1450865623609291904 & 210.37 & 26.77 & $-6.76$ & 1.53 & $-1.27$ & 1.04 & 20.96 & 20.68 & 20.62 & 20.54 \\
1455465494928155008 & 207.19 & 30.12 & $-5.13$ & 1.51 & $-0.05$ & 1.28 & 20.89 & 20.61 & 20.60 & 20.92 \\
4450535168989619968 & 242.70 & 7.60 & $-7.06$ & 0.88 & $-0.23$ & 0.72 & 20.70 & 20.40 & 20.37 & 21.07 \\
4452147808950523520 & 244.09 & 8.40 & $-7.35$ & 0.32 & 0.78 & 0.25 & 19.47 & 19.27 & 19.23 & 21.11 \\
\enddata
\end{deluxetable*}

\begin{figure*}[t]
\centering
\includegraphics[width=1\textwidth]{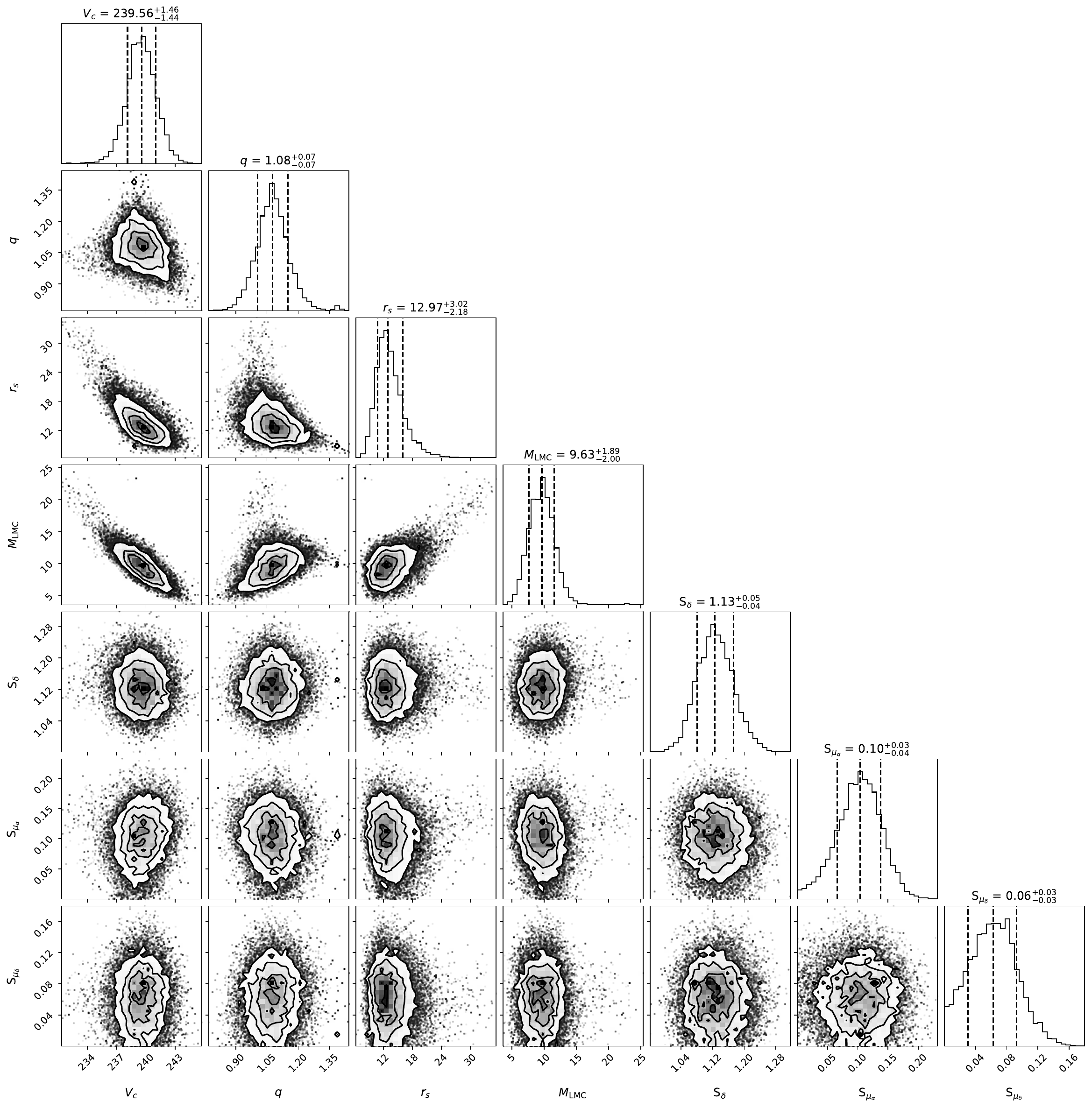}
\caption{Corner plot of the MCMC sampling results for the halo parameters and the mass of the LMC, obtained with the gravitational influence of the LMC included. This plot displays the marginalized posterior distributions and pairwise correlations for the parameters: $V_{\rm c}\ (\mathrm{km\ s^{-1}}),\ q,\ r_s\ (\mathrm{kpc})$, $M_{\rm LMC}$ ($\mathrm{10^{10}\,M_\odot})$, $S_{\delta} (\mathrm{deg})$, $S_{\mu_\alpha} (\mathrm{mas\ yr^{-1}})$, $S_{\mu_\delta} (\mathrm{mas\ yr^{-1}})$. The three vertical dashed lines correspond to the 16th, 50th, and 84th percentiles of each posterior.}
\label{fig:5}
\end{figure*}

\section{Constraining galactic potential using NGC~5466 stream}
\label{sec:4}

Having highly-weighted member candidates of NGC 5466 stream, we aim to investigate the interplay between the stream and the Galactic gravitational field. We construct mock stellar streams and compare them with the observations. We model the disruption of NGC~5466 in a Milky Way potential that includes the gravitational influence of the LMC.

Specifically, we choose to fix the baryonic  components by adopting  the Galactic bulge and disk from \texttt{MWPotential2014} in \texttt{galpy} package \citep{2014ascl.soft11008B}. We adopt an axisymmetric extension of the Navarro--Frenk--White (NFW) profile \citep{1996ApJ...462..563N} to account for deviations from spherical symmetry in the Milky Way halo. The density profile of the halo is given by
\begin{equation}
\rho(m) = \rho_0 \left(\frac{m}{r_s}\right)^{-1} \left[1+\frac{m}{r_s}\right]^{-2},
\label{eq:spheroid_density}
\end{equation}
where $r_s$ is the scale radius of the halo, and $\rho_0$ is the scale density. The spheroidal radius $m$ is defined as
\begin{equation}
m^2 = R^2 + \frac{z^2}{q^2},
\end{equation}
with $R$ being the cylindrical galactocentric radius ($R=\sqrt{x^2+y^2}$), $z$ the height above the Galactic plane, and $q$ the axis ratio of the density contours that describes the shape flattening of the dark halo. 

To isolate the dark matter halo's contribution to the local rotation curve, 
$V_{\rm halo}$, we fix the baryonic velocity component $V_{\rm bary}$ and 
sample the total circular velocity $V_{\rm c}$ such that
\begin{equation}
V_{\rm halo}^2(R_0) = V_{\rm c}^2(R_0) - V_{\rm bary}^2(R_0).
\end{equation}
Subsequently, the scale density $\rho_0$ of the NFW halo is uniquely normalized by ensuring that the theoretical circular velocity of the dark halo at the solar radius $R_0 = 8.122\,{\rm kpc}$ perfectly matches this deduced $V_{\rm halo}$. In other words, we sample the dark halo mass by indirectly sampling $V_{\rm c}^2(R_0)$. For a random $V_{\rm c}^2(R_0)$, to transform coordinates between ICRS and Galactocentric frame, the solar peculiar motion $(U_\odot, V_\odot, W_\odot) = (11.1, 12.24, 7.25)\ {\rm km \, s^{-1}}$ \citep{2010MNRAS.403.1829S}
 is added to $V_{\rm c}^2(R_0)$ to get the Sun's total velocity relative to the Galactic center. Other attributes stay the same as those of ``v4.0'' in \texttt{astropy} .

The setup for the LMC is identical to that in Section~\ref{sec:3.1}, except that the LMC mass is treated as a free parameter, and its scale radius is determined according to Equation~\ref{eq:7}. The mock stream is generated adopting the same approach as detailed in Section~\ref{sec:3.1}. We trace the orbit of NGC~5466 backward for 5~Gyr, and then integrate forward to the present day.

We construct a Bayesian posterior probability function based on both observational and mock data. 
The underlying Galactic gravitational potential is parameterized by a set of core parameters $\boldsymbol{\theta}_{\text{pot}} = \{V_c, q, r_s, M_{\text{LMC}}\}$, where $V_c$ represents the circular velocity at the solar radius, $q$ is the density flattening of halo, $r_s$ is the scale radius of the halo density profile, and $M_{\text{LMC}}$ denotes the total mass of the LMC.

To extract a robust and noise-resistant representation of the modeled stream track from the discrete particle-spray realization, we sort the simulated stream particles by $\alpha$ and divide them into uniform angular bins with equal widths in $\alpha$. Therefore, each bin represents an equal angular interval on the sky rather than an equal number of particles or a uniform surface density. Within each $\alpha$ bin, we calculate the median values of $\delta$ and proper motions ($\mu_{\alpha^*}$ and $\mu_{\delta}$). If an empty bin produces an undefined statistic, one-dimensional linear interpolation along $\alpha$ using adjacent valid bins is applied to construct the smooth model stream track used in the likelihood calculation.

We employ a Markov Chain Monte Carlo (MCMC) Bayesian inference framework to sample the joint posterior probability distribution $P(\boldsymbol{\theta} | \mathbf{D})$ given the full observational dataset $\mathbf{D}$. According to Bayes' theorem, the log-posterior probability satisfies:
\begin{equation}
\ln P(\boldsymbol{\theta} | \mathbf{D}) \propto \ln P(\boldsymbol{\theta}) + \ln \mathcal{L}(\mathbf{D} | \boldsymbol{\theta})
\end{equation}
where $\boldsymbol{\theta} = \{V_c, q, r_s, M_{\text{LMC}}, S_{\delta}, S_{\mu_\alpha}, S_{\mu_\delta}\}$ is the 7D parameter vector. Here $S_{\delta}$, $S_{\mu_\alpha}$, and $S_{\mu_\delta}$ represent the intrinsic physical widths and kinematic dispersion vectors of the stellar stream, acting as vital nuisance parameters to encompass the finite spatial and velocity distribution of the real tidal structure.
We enforce uniform priors on all parameters. The log-prior $\ln P(\boldsymbol{\theta})$ is expressed as:
\begin{equation}
\ln P(\boldsymbol{\theta}) = 
\begin{cases} 
0, & \text{if } \boldsymbol{\theta} \in \mathbf{\Omega} \\
-\infty, & \text{otherwise}
\end{cases}
\end{equation}
where the boundary of the allowed parameter space $\mathbf{\Omega}$ is explicitly defined by:

\begin{equation}
\begin{aligned}
210 < &V_c \ [\mathrm{km\ s^{-1}}] < 250 \\
0.3 < &q < 1.5 \\
5 < &r_s \ [\mathrm{kpc}] < 40 \\
1 < &M_{\text{LMC}} \ [10^{10} \mathrm{M_\odot}] < 30 \\
0.5 < &S_{\delta} \ [\mathrm{deg}] < 2 \\
0 < &S_{\mu_\alpha} \ [\mathrm{mas\ yr^{-1}}] < 0.5 \\
0 < &S_{\mu_\delta} \ [\mathrm{mas\ yr^{-1}}] < 0.5
\end{aligned}
\end{equation}

Assuming the observational errors across individual stars are independent and follow a Gaussian distribution, the total log-likelihood $\ln \mathcal{L}(\mathbf{D} | \boldsymbol{\theta})$ is:
\begin{equation}
\ln \mathcal{L}(\mathbf{D} | \boldsymbol{\theta}) = \ln \mathcal{L}_{\delta} + \ln \mathcal{L}_{\mu_{\alpha^*}} + \ln \mathcal{L}_{\mu_\delta}
\end{equation}

For each coordinate dimension, the component log-likelihood is formalized as:
\begin{multline}
\ln \mathcal{L}_{\text{comp}} = \sum_{j=1}^{N} [ -\frac{1}{2} \ln ( 2\pi \sigma_{\text{tot}, j}^2 ) \\
- \frac{( X_{\text{obs}, j} - X_{\text{model}}(\alpha_j) )^2}{2\sigma_{\text{tot}, j}^2} ]
\end{multline}
where $X_{\text{model}}(\alpha_j)$ represents the predicted path coordinate mapped to the individual star's observed $\alpha_j$ using a linear interpolation mapped from the mock stream.

 The total effective variance $\sigma_{\text{tot}, j}^2$ for the $j$-th observational star is the quadrature sum of its observational measurement error $\sigma_{\text{obs}, j}$ and the structural intrinsic scatter parameter $S_{\text{comp}}$:
\begin{equation}
\sigma_{\text{tot}, j}^2 = \sigma_{\text{obs}, j}^2 + S_{\text{comp}}^2
\end{equation}

We employ 64 walkers in \texttt{emcee} \citep{2013PASP..125..306F} for the sampling process. Each walker undergoes a 2000-step burn-in phase to ensure convergence, followed by 2000 effective sampling steps. Particle-spray stream realizations are generated dynamically at each MCMC step according to the proposed candidate potential parameters.

Figure~\ref{fig:5} shows the Bayesian posterior distributions of the dark matter parameters and the mass of the LMC, where the three vertical dashed lines indicate the 16th, 50th, and 84th percentiles of the posteriors. From the posterior distributions, we obtain the following constrained values: 
$V_{\rm c} = 239.56^{+1.46}_{-1.44}\ \mathrm{km\ s^{-1}}, q = 1.08^{+0.07}_{-0.07}, r_s = 12.97^{+3.02}_{-2.18}\ \mathrm{kpc}, M_{\rm LMC} = 9.63^{+1.89}_{-2.00}\times 10^{10}\ \mathrm{M_\odot}$.

\section{Discussion}
\label{sec:5}
\begin{figure*}[t]
\centering
\includegraphics[width=0.3\textwidth]{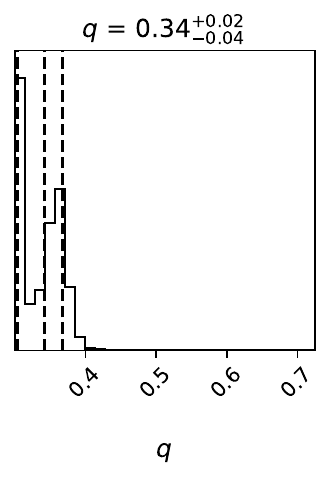}
\caption{Bayesian posterior distribution of the density flattening of halo obtained without including the gravitational influence of the LMC, where the three vertical dashed lines indicate the 16th, 50th, and 84th percentiles of the posteriors.}
\label{fig:6}
\end{figure*}

Besides the modeling in the presence of LMC, we also perform the similar MCMC fitting without LMC, that is, to the Milky Way dark halo alone. Figure~\ref{fig:6} shows the Bayesian posterior distribution of the density flattening of halo obtained without including the gravitational influence of the LMC, where the three vertical dashed lines indicate the 16th, 50th, and 84th percentiles of the posteriors. From the posterior distributions, we obtain the following constrained values: 
$V_{\rm c} = 245.85^{+1.49}_{-1.31}\ \mathrm{km\ s^{-1}}, q = 0.34^{+0.02}_{-0.04}, r_s = 12.86^{+2.35}_{-0.90}\ \mathrm{kpc}$. Compared with the Bayesian posterior distributions derived from models including the influence of the LMC, the density flattening of halo $q$ inferred from models without the LMC become extremely low, which seems unreasonable. This indicates that the inclusion of the LMC provides more effective dynamical constraints on the halo parameters.

NGC 5466 mainly constrains the large-scale response of the Galactic potential and the accumulated orbital perturbation. Therefore, the inferred value should be interpreted as a constraint from a relatively weakly perturbed stream. In a MW-only framework, ignoring the time-dependent perturbation from the massive LMC forces the model to absorb these dynamical effects into the geometric parameters, driving the flattening $q$ to an unphysical value. This highlights that incorporating the LMC potential is indispensable for accurately determining the halo shape.

Our inferred density flattening of the dark matter halo, $q=1.08^{+0.07}_{-0.07}$, indicates a nearly spherical halo. Here, $q$ refers to the axis ratio of the halo density distribution in our axisymmetric NFW model, where $q=1$ corresponds to a spherical halo, $q<1$ represents an oblate density distribution, and $q>1$ represents a prolate density distribution. We note that flattening parameters reported in previous studies may describe either the density distribution or the gravitational potential, and therefore are not always directly comparable to our $q$. Our result is consistent with analyses such as \citet{2023MNRAS.524.2124P}, who also favor a near-spherical halo based on a joint analysis of multiple stellar streams, including NGC~3201, M68, and Palomar~5. They report a density flattening of $q=1.06\pm0.06$, which is directly comparable to our density flattening parameter. By dynamically modeling the phase-space tracks of the Palomar~5 and GD-1 stellar streams, \citet{2016ApJ...833...31B} determined the axis ratio of the dark matter halo density distribution to be $1.05\pm0.14$ within the inner $20\,{\rm kpc}$, also favoring a nearly spherical density profile. \citet{2019sf2a.conf..495W} applied the Jeans equations to a large sample of $\sim$15,600 RR~Lyrae stars from \textit{Gaia}~DR2 to non-parametrically map the Galactic acceleration field. After subtracting the baryonic components, they measured a dark matter density flattening of $q=1.00\pm0.09$ between $5$ and $20\,{\rm kpc}$. Their inferred density flattening is consistent with our result. In contrast, \citet{2025ApJ...985L..22N} reported a $z$-axis density flattening of $q=0.81^{+0.06}_{-0.03}$ from modeling the GD-1 stellar stream under an axisymmetric assumption. However, their analysis further reveals a $2\sigma$ deviation from axisymmetry, preferring a tilted triaxial halo with density axis ratios of $1:0.75:0.70$. Overall, our inferred $q$ supports a near-spherical Milky Way dark matter halo within an axisymmetric framework and remains broadly consistent with contemporary constraints from stellar streams.

\begin{figure*}[t]
\centering
\includegraphics[width=1\textwidth]{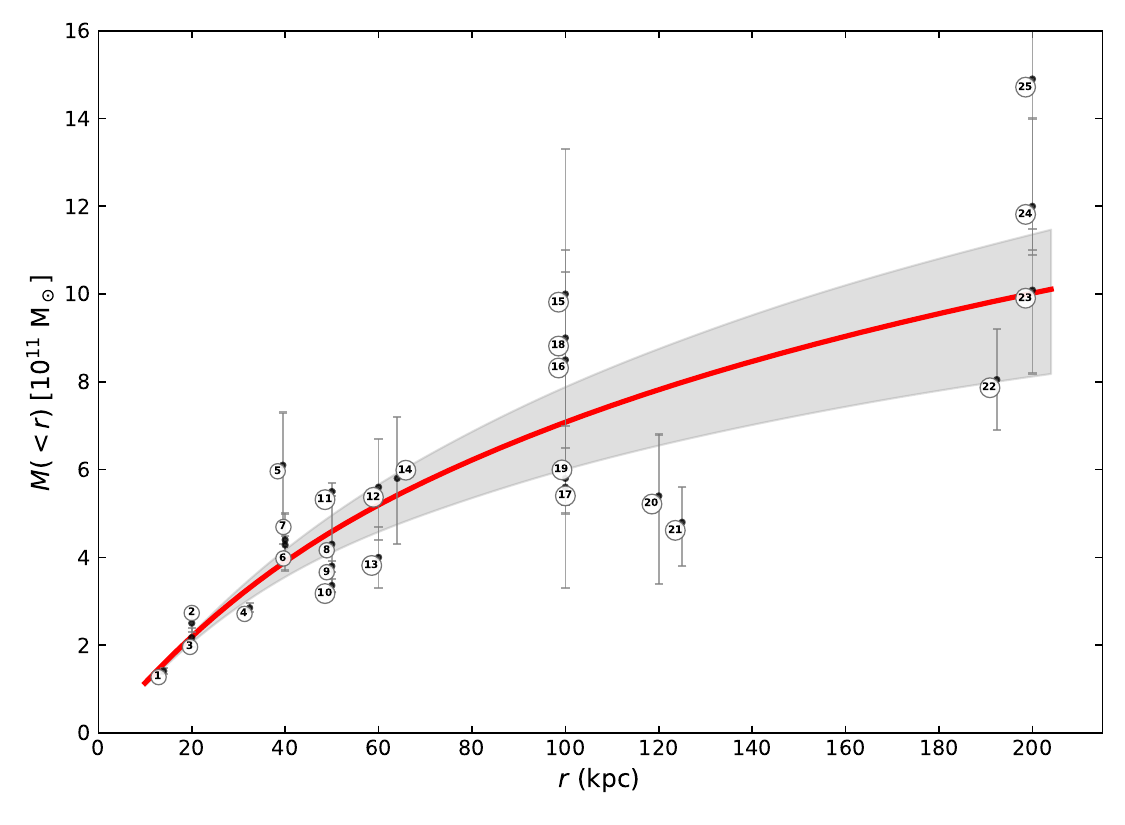}
\caption{Comparison of the Milky Way enclosed mass measurements, showing posterior medians and 1$\sigma$ uncertainties from this work and previous studies. The numbers correspond to the following studies:
(1) \citet{2025ApJ...985L..22N}; 
(2) \citet{2019MNRAS.486.2995M}; 
(3) \citet{2025MNRAS.541..214D}; 
(4) \citet{2023MNRAS.521.4936K}; 
(5) \citet{2018ApJ...862...52S}; 
(6) \citet{2025ApJ...987..143Z}; 
(7) \citet{2019ApJ...873..118W}; 
(8) \citet{2015ApJ...803...80K} at $50\,\mathrm{kpc}$; 
(9) \citet{2019MNRAS.487.2685E}; 
(10) \citet{2026MNRAS.tmp..856B}; 
(11) \citet{2003A&A...397..899S}; 
(12) \citet{2018MNRAS.479..570H}; 
(13) \citet{2008ApJ...684.1143X}; 
(14) \citet{2020MNRAS.494.5178F}; 
(15) \citet{2022MNRAS.516..731B}; 
(16) \citet{2019MNRAS.484.2832V}; 
(17) \citet{2021MNRAS.501.2279V}; 
(18) \citet{2015ApJ...803...80K} at $100\,\mathrm{kpc}$; 
(19) \citet{2025arXiv250819351M}; 
(20) \citet{2005MNRAS.364..433B}; 
(21) \citet{2017ApJ...835..167E}; 
(22) \citet{2023ApJ...946...73Z}; 
(23) \citet{2024ApJ...967...89I}; 
(24) \citet{2013A&A...549A.137I}; 
(25) \citet{2025MNRAS.538.1442L}. 
}
\label{fig:7}
\end{figure*}

Figure~\ref{fig:7} shows the comparison of the Milky Way enclosed mass measurements, showing posterior medians and 1$\sigma$ uncertainties from this work and previous studies. Based on our modeling of the NGC~5466 tidal stream, we estimate the enclosed mass of the Milky Way within the stream apocenter radius of $\sim$ 50 kpc to be $M_{\rm MW}(<50~\mathrm{kpc}) = 4.59^{+0.48}_{-0.37} \times 10^{11}$ $\mathrm{M_\odot}$. In closer alignment with our single-stream modeling approach, \citet{2015ApJ...803...80K} analyzed the tidal stream of the globular cluster Palomar~5 within a Bayesian framework based on streakline stream models and a continuous likelihood function, constraining the enclosed mass of the Milky Way to be $M_{\rm MW}(<50~\mathrm{kpc}) = 4.3 \pm 1.1 \times 10^{11}$ $\mathrm{M_\odot}$. \citet{2024ApJ...967...89I} selected a comprehensive sample of 29 long and thin cold tidal streams detected in \textit{Gaia}~DR3 using the \texttt{STREAMFINDER} algorithm. By modeling the coherent phase-space trajectories of these 29 streams within a flexible Galactic potential via MCMC fitting, they constrained the enclosed Milky Way mass within $50\,{\rm kpc}$ to be $M_{\rm MW}(<50\,{\rm kpc}) = (4.6 \pm 0.3) \times 10^{11}\,\mathrm{M_\odot}$. Meanwhile, adopting a different tracer population, \citet{2026MNRAS.tmp..856B} used outer halo stellar data out to $160\ \mathrm{kpc}$ from the combined H3+SEGUE+MagE survey and constrained the enclosed Milky Way mass to be $M_{\rm MW}(<50~\mathrm{kpc}) = 3.36 \pm 0.15 \times 10^{11}$ $\mathrm{M_\odot}$.  \citet{2019MNRAS.487.2685E} utilized the velocity-track misalignment of the Orphan stream induced by the LMC to simultaneously fit the host and satellite potentials, yielding an enclosed Milky Way mass of $M_{\rm MW}(<50\,{\rm kpc}) = 3.80^{+0.14}_{-0.11} \times 10^{11}\,{\rm M_\odot}$.

\citet{2019MNRAS.487.2685E} found that a total LMC mass of $1.38 ^{+0.27}_{-0.24}\times 10^{11}\,\mathrm{M_\odot}$ was necessary to explain the observed bending of the Orphan stream. \citet{2023MNRAS.521.4936K} used the 6D structure of the Orphan–Chenab stream to precisely constrain the LMC mass profile, measuring an enclosed mass of $M_{\rm LMC}(<32.8\,{\rm kpc}) = 7.02 \pm 0.90\times 10^{10}\,\mathrm{M_\odot}$, and suggested the LMC’s total mass was $1.29 ^{+0.28}_{-0.23}\times 10^{11}\,\mathrm{M_\odot}$. \citet{2026MNRAS.tmp..856B}, using outer-halo kinematics combined with simulation-based inference, inferred a LMC mass of $M_{\rm LMC}(<50\,{\rm kpc}) = 8.76^{+1.94}_{-1.77}\times 10^{10}\,\mathrm{M_\odot}$. \citet{2021MNRAS.501.2279V} used the Sagittarius stream to constrain the LMC mass, obtaining a total mass of approximately $1.3 \pm 0.3\times 10^{11}\,\mathrm{M_\odot}$. \citet{2021ApJ...923..149S} used the Orphan--Chenab stream to constrain the LMC mass, obtaining a total mass of $1.88^{+0.35}_{-0.40} \times 10^{11}\,\mathrm{M_\odot}$. Our estimate of $M_{\rm LMC} = 9.63^{+1.89}_{-2.00}\times 10^{10}\ \mathrm{M_\odot}$ is somewhat lower, which may reflect the limited sensitivity of NGC~5466 to the LMC, since the stream just exhibits a relatively small portion ($180^\circ<\alpha<200^\circ$) that is heavily affected by the LMC perturbation. As a northern stellar stream, NGC~5466 may nevertheless provide a complementary probe of the LMC potential when combined with other streams.

\section{Conclusions}
\label{sec:6}

In this work, we analyze the tidal stream of NGC~5466 by combining photometric data from LS DR9 and PS DR2 with astrometric data from \textit{Gaia} DR3. By applying a modified matched filter method that incorporates both color--magnitude and proper-motion information, we identified a high-quality sample of 308 stream member candidates. An inspection of the weighted sky map reveals that the tidal stream of NGC~5466 extends over $80^\circ$ on the sky. The stream can be continuously traced across this angular range, indicating that the tidal stream of NGC~5466 is more extended than previously reported. We also detect the leading end towards $\alpha<190^\circ$ that was thought to be ambiguous and deviated from the main stream track.

Using these member candidates, we constructed mock streams under a range of Milky Way halo potentials and employed a Bayesian framework to constrain the Galactic dark matter halo parameters. Our analysis yields a circular velocity at the solar radius of $V_{\rm c} = 239.56^{+1.46}_{-1.44}\ \mathrm{km\ s^{-1}}$, a nearly spherical dark matter halo of $q = 1.08^{+0.07}_{-0.07}$, and a scale radius of $ r_s = 12.97^{+3.02}_{-2.18}\ \mathrm{kpc}$. We estimate the enclosed mass of the Milky Way within the stream apocenter radius of $\sim$ 50 kpc to be $M_{\rm MW}(<50~\mathrm{kpc}) = 4.59^{+0.48}_{-0.37} \times 10^{11}$ $\mathrm{M_\odot}$. Using the deviation part at the lead end of the stream, we also estimate the total mass of the LMC to be $M_{\rm LMC} = 9.63^{+1.89}_{-2.00}\times 10^{10}\,\mathrm{M_\odot}$.

\section*{Acknowledgements}

%We thank the referee for the thorough reviews that helped us improve the manuscript. 

This study is supported by the National Key R\&D Program of China, grant nos. 2024YFA1611902 and 2023YFE0107800, the National Natural Science Foundation of China under grant nos. 12588202, 12273055, and 12503024,  the International Partnership Program of Chinese Academy of Sciences under grant No. 178GJHZ2022040GC, the support from the Strategic Priority Research Program of Chinese Academy of Sciences grant No. XDB1160301 and the support by the National Astronomical Observatories, CAS grant no. E4ZB0301.

The Legacy Surveys consist of three individual and complementary projects: the Dark Energy Camera Legacy Survey (DECaLS; Proposal ID \#2014B-0404; PIs: David Schlegel and Arjun Dey), the Beijing-Arizona Sky Survey (BASS; NOAO Prop. ID \#2015A-0801; PIs: Zhou Xu and Xiaohui Fan), and the Mayall z-band Legacy Survey (MzLS; Prop. ID \#2016A-0453; PI: Arjun Dey). DECaLS, BASS and MzLS together include data obtained, respectively, at the Blanco telescope, Cerro Tololo Inter-American Observatory, NSF’s NOIRLab; the Bok telescope, Steward Observatory, University of Arizona; and the Mayall telescope, Kitt Peak National Observatory, NOIRLab. Pipeline processing and analyses of the data were supported by NOIRLab and the Lawrence Berkeley National Laboratory (LBNL). The Legacy Surveys project is honored to be permitted to conduct astronomical research on Iolkam Du’ag (Kitt Peak), a mountain with particular significance to the Tohono O’odham Nation.

NOIRLab is operated by the Association of Universities for Research in Astronomy (AURA) under a cooperative agreement with the National Science Foundation. LBNL is managed by the Regents of the University of California under contract to the U.S. Department of Energy.

This project used data obtained with the Dark Energy Camera (DECam), which was constructed by the Dark Energy Survey (DES) collaboration. Funding for the DES Projects has been provided by the U.S. Department of Energy, the U.S. National Science Foundation, the Ministry of Science and Education of Spain, the Science and Technology Facilities Council of the United Kingdom, the Higher Education Funding Council for England, the National Center for Supercomputing Applications at the University of Illinois at Urbana-Champaign, the Kavli Institute of Cosmological Physics at the University of Chicago, Center for Cosmology and Astro-Particle Physics at the Ohio State University, the Mitchell Institute for Fundamental Physics and Astronomy at Texas A\&M University, Financiadora de Estudos e Projetos, Fundacao Carlos Chagas Filho de Amparo, Financiadora de Estudos e Projetos, Fundacao Carlos Chagas Filho de Amparo a Pesquisa do Estado do Rio de Janeiro, Conselho Nacional de Desenvolvimento Cientifico e Tecnologico and the Ministerio da Ciencia, Tecnologia e Inovacao, the Deutsche Forschungsgemeinschaft and the Collaborating Institutions in the Dark Energy Survey. The Collaborating Institutions are Argonne National Laboratory, the University of California at Santa Cruz, the University of Cambridge, Centro de Investigaciones Energeticas, Medioambientales y Tecnologicas-Madrid, the University of Chicago, University College London, the DES-Brazil Consortium, the University of Edinburgh, the Eidgenossische Technische Hochschule (ETH) Zurich, Fermi National Accelerator Laboratory, the University of Illinois at Urbana-Champaign, the Institut de Ciencies de l’Espai (IEEC/CSIC), the Institut de Fisica d’Altes Energies, Lawrence Berkeley National Laboratory, the Ludwig Maximilians Universitat Munchen and the associated Excellence Cluster Universe, the University of Michigan, NSF’s NOIRLab, the University of Nottingham, the Ohio State University, the University of Pennsylvania, the University of Portsmouth, SLAC National Accelerator Laboratory, Stanford University, the University of Sussex, and Texas A\&M University.

BASS is a key project of the Telescope Access Program (TAP), which has been funded by the National Astronomical Observatories of China, the Chinese Academy of Sciences (the Strategic Priority Research Program “The Emergence of Cosmological Structures” Grant \# XDB09000000), and the Special Fund for Astronomy from the Ministry of Finance. The BASS is also supported by the External Cooperation Program of Chinese Academy of Sciences (Grant \# 114A11KYSB20160057), and Chinese National Natural Science Foundation (Grant \# 12120101003, \# 11433005).

The Legacy Survey team makes use of data products from the Near-Earth Object Wide-field Infrared Survey Explorer (NEOWISE), which is a project of the Jet Propulsion Laboratory/California Institute of Technology. NEOWISE is funded by the National Aeronautics and Space Administration.

The Legacy Surveys imaging of the DESI footprint is supported by the Director, Office of Science, Office of High Energy Physics of the U.S. Department of Energy under Contract No. DE-AC02-05CH1123, by the National Energy Research Scientific Computing Center, a DOE Office of Science User Facility under the same contract; and by the U.S. National Science Foundation, Division of Astronomical Sciences under Contract No. AST-0950945 to NOAO.

This research has made use of data from the Pan-STARRS1 Surveys (PS1) and the PS1 public science archive. 
PS1 has been made possible through contributions by the Institute for Astronomy, the University of Hawaii, 
the Pan-STARRS Project Office, the Max-Planck Society and its participating institutes (the Max Planck Institute for Astronomy, Heidelberg, and the Max Planck Institute for Extraterrestrial Physics, Garching), 
The Johns Hopkins University, Durham University, the University of Edinburgh, Queen's University Belfast, 
the Harvard-Smithsonian Center for Astrophysics, the Las Cumbres Observatory Global Telescope Network Incorporated, 
the National Central University of Taiwan, the Space Telescope Science Institute, NASA under Grant No. NNX08AR22G, 
the National Science Foundation Grant No. AST-1238877, the University of Maryland, Eotvos Lorand University (ELTE), 
the Los Alamos National Laboratory, and the Gordon and Betty Moore Foundation.

We also acknowledge the following publications describing the PS1 instrument, survey, and data processing: 
\citet{2016AAS...22732407C}; \citet{2020ApJS..251....3M};
\citet{2020ApJS..251....4W}; 
\citet{2020ApJS..251....5M}; 
\citet{2020ApJS..251....6M}; 
\citet{2020ApJS..251....7F}.

This work presents results from the European Space Agency (ESA) space mission \emph{Gaia}. 
\textit{Gaia} data are processed by the \textit{Gaia} Data Processing and Analysis Consortium (DPAC). Funding for the DPAC is provided by national institutions, in particular the institutions participating in the \textit{Gaia} Multilateral Agreement (MLA). 
The \textit{Gaia} mission website is \url{https://www.cosmos.esa.int/gaia}, and the \textit{Gaia} archive website is \url{https://archives.esac.esa.int/gaia}.

%% For this sample we use BibTeX plus aasjournalv7.bst to generate the

\bibliography{references}{}
\bibliographystyle{aasjournalv7}



\end{document}